\documentclass[twocolumn,a4paper,accepted=2023-08-04]{quantumarticle}%
\pdfoutput=1
\usepackage[T1]{fontenc}
\usepackage[latin9]{inputenc}
\usepackage{geometry}
\usepackage{amsmath}
\usepackage{amssymb}
\usepackage{graphicx}
\usepackage{adjustbox} 
\usepackage[toc,page]{appendix}
\usepackage{xcolor}
\usepackage{hyperref}
\usepackage[numbers]{natbib}
\makeatletter

\providecommand{\tabularnewline}{\\}

\usepackage{babel}

\newcommand{\eq}[1]{Eq.~(\ref{#1})} %
\def\be{\begin{equation}} %
\def\ee{\end{equation}} %
\newcommand{\bea}{\begin{eqnarray}}
\newcommand{\eea}{\end{eqnarray}}

\newcommand{\HU}{\hat U}
\newcommand{\HV}{\hat V}

\newcommand{\HH}{\hat H}

\newcommand{\HI}{\hat I}
\newcommand{\hn}{\hat n}
\newcommand{\hr}{\hat r}

\newcommand{\CR}[1]{\hat a^{\dagger}_{#1}}
\newcommand{\AN}[1]{\hat a_{#1}}

\usepackage{babel}

\makeatother

\usepackage{babel}
\begin{document}
\title{Assessment of various Hamiltonian partitionings for the electronic structure problem on a quantum computer using the Trotter approximation}
\author{Luis A. Mart\'inez-Mart\'inez}
\affiliation{Department of Physical and Environmental Sciences, University of Toronto Scarborough, Toronto, Ontario M1C 1A4, Canada}
\affiliation{Chemical Physics Theory Group, Department of Chemistry, University of Toronto, Toronto, Ontario M5S 3H6, Canada}
\author{Tzu-Ching Yen}
\affiliation{Department of Physical and Environmental Sciences, University of Toronto Scarborough, Toronto, Ontario M1C 1A4, Canada}
\affiliation{Chemical Physics Theory Group, Department of Chemistry, University of Toronto, Toronto, Ontario M5S 3H6, Canada}
\author{Artur F. Izmaylov}
\affiliation{Department of Physical and Environmental Sciences, University of Toronto Scarborough, Toronto, Ontario M1C 1A4, Canada}
\affiliation{Chemical Physics Theory Group, Department of Chemistry, University of Toronto, Toronto, Ontario M5S 3H6, Canada}
\begin{abstract}
Solving the electronic structure problem via unitary evolution of the electronic Hamiltonian is one of 
the promising applications of digital quantum computers. 
One of the practical strategies to implement the unitary evolution is 
via Trotterization, where a sequence of short-time evolutions of 
fast-forwardable (i.e. efficiently diagonalizable) Hamiltonian fragments is used.  
Given multiple choices of possible Hamiltonian decompositions to fast-forwardable fragments, the accuracy
of the Hamiltonian evolution depends on the choice of the fragments.  
We assess efficiency of multiple Hamiltonian partitioning techniques using fermionic and qubit algebras for the Trotterization. 
Use of symmetries of the electronic Hamiltonian and its fragments significantly reduces the Trotter error. This reduction
makes fermionic-based partitioning Trotter errors lower compared to those in qubit-based techniques. 
However, from the simulation-cost standpoint, fermionic methods tend to introduce quantum circuits with a greater number of 
T-gates at each Trotter step and thus are more computationally expensive compared to their qubit counterparts.
\end{abstract}
\maketitle

\section{Introduction}

Solving the electronic structure problem using controlled time evolution of quantum systems is one 
of the most promising applications of digital quantum computers. Ease of mapping fermionic operators 
to qubits using one of few existent mapping makes the second quantized formalism a convenient 
starting point.  The molecular electronic Hamiltonian with $N$ single-particle spin-orbitals is
\begin{equation}
\HH=\sum_{pq=1}^{N}h_{pq}\CR{p}\AN{q}+\sum_{pqrs=1}^{N}g_{pqrs}\CR{p}\AN{q}\CR{r}\AN{s}\label{eq:El_Ham-1}
\end{equation}
where $\CR{p}$ ($\AN{p}$) is the creation (annihilation)
fermionic operator for the $p$th spin-orbital, $h_{pq}$ and $g_{pqrs}$ are one- and two-electron integrals \cite{helgaker2014}.
To obtain the spectrum of $H$ using digital quantum computers with error-correcting capability 
one can use the quantum phase estimation (QPE) algorithm. QPE involves unitary evolution generated by the system Hamiltonian $\exp[-it\HH]$ to obtain the auto-correlation function 
and then Fourier transform (we will use atomic units throughout this work, $\hbar=1$). 

One of the challenges of QPE is that $\exp[-it\HH]$ cannot be implemented straightforwardly
on a digital quantum computer.  Within a fault-tolerant paradigm, several approaches have been put forward to address this problem: 
oracle query-based algorithms \cite{childs2003exponential}, quantum walks \cite{childs2010}, 
qubitization \cite{low2019}, linear combination of unitaries \cite{ChildsLCU} and product formulas \cite{lloyd1996}.
We will refer to the last approach as Trotterization throughout this work. 
Trotterization starts with partitioning the Hamiltonian 
to exactly solvable or fast-forwardable parts, $\HH_n$'s:
\begin{equation}
\HH=\sum_{n=1}^{\Gamma} \HH_n\label{Partition}.
\end{equation}

Using the Trotter approximation, $\exp[-it\HH]$ can be expressed as a sequence of 
$e^{-i\HH_nt}$ transformations for small $t$
\bea\label{eq:1stOT}
e^{-it\HH}=\prod_{n=1}^\Gamma e^{-it\HH_n}+\mathcal{O}(t^{2}),
\eea
where time propagator operators $\exp[-it\HH_n]$ can be translated to quantum gates easily. Implementation 
of $\exp[-it\HH_n]$ propagators is straightforward because there are known unitary transformations 
$\hat V_n$ that bring $\HH_n$ to operators diagonal in computational basis $\hat D_n$, 
$\HH_n = \hat V_n^\dagger \hat D_n \hat V_n$. Due to non-commutativity of individual 
$\HH_n$'s the Trotter error is proportional to commutator norms $||[\HH_n,\HH_{n'}]||$. 

Even though Trotterization exhibits a steeper scaling of quantum circuit resources with target error 
compared to the rest of simulation approaches, it has two unique advantages 
compared to the so-called post-Trotter algorithms \cite{childs2019nearly}: 
1) lower overhead and the number of required ancilla qubits,  
and 2) the unique error scaling with the commutativity 
between the fast-forwardable Hamiltonian fragments, $\{\HH_n\}$. 
The second property can be exploited in the simulation of physical systems where locality 
of interactions can be leveraged to reduce the number of non-zero commutators between Hamiltonian fragments. 

Thus, Trotterized Hamiltonian simulation has been the subject of 
active research aiming at the reduction of the complexity of its associated quantum circuits
in different physical systems \cite{babbush2015chemical,kivlichan2020,tran2020destructive,childs2021Trot,keever2023}.\par
There are several approaches to partitioning of molecular electronic Hamiltonians into exactly solvable fragments. 
Many of these approaches were developed in the measurement problem
\cite{kandala2017,matsuzawa2020,verteletskyi2020,yen2020measuring,motta2021,yen2021} of the 
Variational Quantum Eigensolver (VQE) \cite{peruzzo2014}, where the partitioning is needed to 
obtain the expectation value of the Hamiltonian. A few of these partitioning techniques were considered 
for the Troterization problem \cite{van2020,motta2021}, yet different 
partitioning methods have not been compared systematically. 

In this work, we consider Hamiltonian partitioning methods within two Hamiltonian realizations, which use fermionic and 
qubit operators, respectively. Although both realizations employ exactly solvable Hamiltonians, the structure of these 
Hamiltonians is different. In the fermionic case, the origin of solvability 
can be traced to the well-known solvability of hermitian one-electron Hamiltonians, while in the qubit case, the solvability 
is based on existence of Clifford group transformations that diagonalize any linear combination of commuting 
Pauli products. 

For a Trotterization algorithm, the exponentials in approximate time evolution operator introduced in Eq. (\ref{eq:1stOT}) are decomposed in terms of single-qubit rotations and Clifford gates. 
In an error-corrected algorithm such as surface code, Clifford gates can be implemented fault-tolerantly. 
Single-qubit-rotations, on the other hand, need to be compiled as a series of gates consisting of Clifford 
operations and at least one non-Clifford one, usually taken to be the T gate, a rotation around 
the $z$-axis by $\frac{\pi}{8}$ \cite{nielsen00}. The implementation of T gates requires a procedure called 
magic state distillation \cite{gidney2019}, which renders T-gate synthesis the most costly aspect of error correction. The number of T gates required for a quantum algorithm is one of the essential efficiency determining parameters in fault-tolerant quantum computations. Thus, to determine which Hamiltonian partitioning approach is better, it is reasonable to compare the number of T gates required for performing the approximate unitary evolution. 

The rest of the article is organized as follows. In section \ref{Theory} we introduce the techniques used to analyze our Trotter error results for the different molecular systems and methods. Moreover, the main fermionic and qubit Hamiltonian partition schemes employed in our analysis are introduced. In the same section, we discuss the use of symmetries present in all Hamiltonian fragments that allow us to find tighter estimations of the Trotter error. Section \ref{Results} presents and discusses Trotter errors and T-gate estimates for different partitioning methods. Finally, Section \ref{Conclusions} concludes by summarizing
our most relevant results and giving the outlook.

\section{Theory}
\label{Theory}

\subsection{Upper bounds for Trotter error}
\label{TrotterErrorEst}

Equation \eqref{eq:1stOT} is a first order Trotter-Suzuki formula. Higher order Trotter approximations 
exist for $\exp[-it\HH]$, which feature an error with steeper scaling with $t$, at
the cost of increasing the number of products of exponentials \cite{suzuki1991}.
For simplicity in our analysis, we focus on the error associated to
the first-order Trotter formula. Here, the Trotter error can be written in terms of the spectral norm 
of the difference \cite{childs2021Trot} 
\begin{equation}
||e^{-it\HH}-\prod_n e^{-it\HH_n}|| \le \bar{\alpha} t^{2}/2,\label{OrderErr}
\end{equation}
where
\begin{equation}
\bar{\alpha} =\sum_n ||[\HH_{n},\sum_{n'< n} \HH_{n'}]||\label{alphaP}.
\end{equation}

To avoid order dependent parameter $\bar{\alpha}$ one can use the triangle inequality to upper bound it with  
a simpler quantity 
\bea
2 \sum_n ||[\HH_{n},\sum_{n'< n} \HH_{n'}]|| \le \sum_{n,n'} ||[\HH_{n}, \HH_{n'}]|| = \alpha.
\eea

For Trotter error analysis it is convenient to derive an upper bound on the commutator 
norm ($\alpha$) that contains properties of individual fragments.  
Such an upper bound can provide an insight on what properties of fragments define the success 
of a partitioning in lowering $\alpha$.  

Using the triangle inequality
\bea
|| [\HH_n, \HH_k]|| \le 2 ||\HH_n|| \cdot ||\HH_k||
\eea
provides a loose upper bound. This bound can be further tightened by considering shifted Hamiltonians
\begin{equation}
\overline{\HH}_{i}=\HH_{i}-s_{i}\mathbf{\HI},\label{eq:ShiftOp}
\end{equation}
where $s_{i}$ is a scalar, and $\mathbf{\HI}$ is the identity
operator. The shift introduced in Eq. (\ref{eq:ShiftOp}) leaves the
commutators between Hamiltonian fragments invariant 
\begin{equation}
[\HH_{i},\HH_{j}]=[\overline{\HH}_{i},\overline{\HH}_{j}]\label{Comm_Inv}
\end{equation}
but changes the spectral norms, $||\overline{\HH}_{i}||\neq||\HH_{i}||$
by shifting the spectrum of $\HH_{i}$ in \eq{eq:ShiftOp}. A tighter upper bound for $\alpha $ can be written as 
\bea
\alpha  &=& 2\sum_{i>j}||[\overline{\HH}_{i},\overline{\HH}_{j}]||\le 
4\sum_{i>j}||\overline{\HH}_{i}||||\overline{\HH}_{j}|| \\
&\le& 4\sum_{i>j} \min_{s_i,s_j} ||\overline{\HH}_{i}||\cdot ||\overline{\HH}_{j}||.
\label{TightBet}
\eea
Finding the minimum norm shifts can be done explicitly as
\bea\label{eq:dEdef}
\min_{s_{i}}||\HH_{i}-s_{i}\mathbf{I}||=\frac{\Delta E_{\text{i}}}{2},
\eea
where 
\begin{equation}
\Delta E_{k}=\left(E_{\max,\text{k}}-E_{\min,k}\right)\label{Spec_Range}
\end{equation}
is the spectral range for the $k$th Hamiltonian fragment and $E_{\max,k}$
($E_{\min,k}$) is the maximum (minimum) eigenvalue of
$H_{k}$. Then 
\begin{equation}
\alpha\le \sum_{i>j}\Delta E_{i}\Delta E_{j}=\beta ,\label{Beta_P}
\end{equation}
constitutes the tightest upper bound for $\alpha$ among those involving triangle inequality for shifted Hamiltonian fragments, and it involves only spectral properties of individual fragments.

$\beta$ upper bound can be further analyzed using statistical quantities of fragment's spectral ranges 
\bea
\beta&=& \frac{1}{2}\left[\left(\sum_{i}\Delta E_i\right)^{2}-\sum_{i}\left(\Delta E_i\right)^{2}\right]\\ 
\label{eq:beta}
 & =& \frac{1}{2}  \left(\sum_{i}\Delta E_i\right)^{2} \left( 1 - \sum_i \omega_i^2 \right),
 \eea
 where positive weights $\omega_i$ are defined as 
 \bea
 \omega_i &=& \frac{\Delta E_i}{\sum_k \Delta E_k}, \quad \omega_i\in [0,1]. 
 \eea
 
 Introducing weights $\omega_i$ allows us to recognize in the second factor of \eq{eq:beta}
 the linearized entropy 
 \bea
 S_L = 1 - \sum_i \omega_i^2.\label{LinEntropy}
 \eea
 Therefore, by denoting the first factor in \eq{eq:beta} as $C = \sum_{i}\Delta E_i$ we obtain 
 \bea
 \beta & =&\frac{1}{2} C^{2} S_L.\label{CandS}
\label{Scal_Var}
\eea

Reducing both $C$ and $S_L$ will decrease $\beta$. For reducing $C$ one needs to reduce 
$\Delta E_i$ of the fragments. $S_L$ can be decreased by selecting fragments 
with unevenly distributed weights. 

Finding spectral ranges for exactly solvable problems is not always straightforward because one needs to find 
the largest and lowest eigenvalues among exponentially many. There exists an upper bound for 
$\Delta E/2$ that is based on $L_1$ norm of a coefficient vector for a linear combination of unitaries (LCU) decomposition for the corresponding operator \cite{Ignacio_LCU:2022}. Let us consider an LCU for $\HH_i$  
\bea
\HH_i = \sum_k c_k^{(i)} \HV_k + d_i\mathbf{\HI},
\eea
where $\HV_k$ are some unitaries, $c_k^{(i)}$ and $d_i$ are coefficients. Then, there is a following sequence of inequalities
\bea\label{eq:L1UB}
\frac{\Delta E_i}{2} \le || \HH_i - d_i\mathbf{\HI} || \le \sum_k |c_k^{(i)}|,
\eea
where we used \eq{eq:dEdef} in the first inequality and the triangle inequality with accounting for $||\HV_k||=1$ 
in the second one.  This upper bound suggests that fragments with lower LCU decomposition 
$L_1$ norms can be better candidates for reducing the Trotter error.   

\subsection{T-gate estimates}

Time-energy uncertainty principle requires propagation for longer time to reach higher accuracy in the energy estimation.
The Trotter error bound in \eq{OrderErr} for a single Trotter step can be extended to multiple steps, $N_s$
\bea
\epsilon=||e^{-iT\HH}-\left(\prod_n e^{-it\HH_n}\right)^{N_s}||\le\frac{\alpha  T^{2}}{N_s},
\eea
where $T=N_s t$ is the simulation time. Here, 
we used the fact that the overall error from repeating the first-order Trotter sequence
 $N_s$ times accumulates at most linearly with $N_s$ \cite{Layden}. The Trotter error $\alpha$ is an important figure of merit for the estimation of quantum resources. The number of Trotter steps to reach the level
of accuracy $\epsilon$ while keeping $T$ fixed is $N_s\le\alpha  T^{2}/\epsilon$, so lowering 
$\alpha$ would allow to take longer and fewer time-steps to cover $T$. 

The number of T-gates will be proportional to $N_s$ multiplied by the number of T-gates required for each 
time-step. For comparison of different partitioning methods we will only use system independent 
characteristics, such as the product $\alpha N_R$, where $N_{R}=\sum_{i=1}^{\Gamma}N_{R}^{(i)}$ 
is the number of single-qubit rotations needed in each Trotter step obtained as a sum over 
the number of single-qubit rotations for each Hamiltonian fragment $N_{R}^{(i)}$.
The number of single-qubit rotations with arbitrary angles is proportional to the number of T gates, where the 
coefficient of proportionality can depend on a particular compiling scheme \cite{gheorghiu2022, Mukhopadhyay2022}.

In Appendix \ref{NumTgat} we develop an analysis of the T-gate count for Adaptive QPE eigenvalue estimation for a fixed target error, considering relevant sources of error in addition to the Trotter approximation. This was done with the purpose of showing that the figure of merit $\alpha N_{R}$ suffices to compare the resource-efficiency of the different Hamiltonian decomposition methods addressed in this work.

\subsection{Fermionic partitioning methods}\label{Ferm_Meths}

These methods are based on solvability of one-electron Hamiltonians using orbital rotations, for a one-electron part of the electronic Hamiltonian we can write 
\bea
\hat h_{\rm 1e} &=& \sum_{pq}h_{pq}\CR{p}\AN{q} = \HU^\dagger \left( \sum_p \epsilon^{(1)}_p \hn_p \right) \HU,
\eea
where $\hn_p = \CR{p}\AN{p}$ occupation number operators, $\epsilon_p$ are real constants, and $\HU$ are orbital rotation transformations
\bea\label{eq:Uor}
\HU = \prod_{p>q} e^{\theta_{pq} (\CR{p}\AN{q} - \CR{q}\AN{p})}.
\eea  

Note that $\hn_p$ form a largest commuting subset of $\{\CR{p}\AN{q}\}$ operators, $[\hn_p,\hn_q]=0$, 
and they are directly mapped to polynomial functions of Pauli-$\hat z$ operators by all standard fermion-qubit  mappings \cite{seeley2012}. Orbital rotations
$\HU$ form a continuous Lie group, $\HU_1\HU_2= \HU_3$, therefore, even though different exponential operators  in \eq{eq:Uor} do not commute, one can write each $\HU$ using different orders. This will only require adjusting corresponding amplitudes $\theta_{pq}$. 

Two-electron Hamiltonians that are exact squares of one-electron Hamiltonians also can be solved by 
orbital rotations
\bea
\HH^{(LR)}&=&  \left(\sum_{pq}{\tilde h}_{pq}\CR{p}\AN{q}\right)^2 = 
\HU^\dagger \left( \sum_p {\tilde\epsilon}_p \hn_p \right)^2 \HU \\ \label{eq:Hs}
&=& \HU^\dagger \left( \sum_{p,q} {\tilde \epsilon}_p {\tilde \epsilon}_q \hn_p \hn_q \right) \HU.
\eea 

Formally, ${\tilde \epsilon}_p {\tilde \epsilon}_q$ can be considered as an entry of a rank-deficient matrix, this consideration 
suggests that one can substitute it in principle by a full-rank hermitian matrix, $\lambda_{pq}$, without loss of solvability. Thus a more general two-electron Hamiltonian that can be solved by orbital rotations has a form
\bea\label{eq:Hmf}
\HH^{\rm (FR)} = \HU^\dagger \left( \sum_{p,q} \lambda_{pq} \hn_p \hn_q \right) \HU.
\eea  

Note that if $\lambda_{pq}$ is a diagonal matrix $\HH^{\rm (FR)}$ can describe purely one-electron Hamiltonians,
this is in contrast with a rank-deficient case where ${\tilde \epsilon}_p {\tilde \epsilon}_q$ 
can encode purely one-electron 
Hamiltonians if ${\tilde \epsilon}_p$ has only a single non-zero component. 
Further generalization of two-electron Hamiltonians solvable by orbital rotations is done in Ref. \cite{Izmaylov:MF}, 
but it follows the idea of having different eigenstates obtained by 
different orbital rotations and is not as straightforward to use as forms in \eq{eq:Hmf} and \eq{eq:Hs}. 
  
\emph{Low-rank (LR) decomposition:} This method uses orbital rotations to diagonalize the 
one-electron part and to represent the two-electron part of \eq{eq:El_Ham-1} as a linear combination of \eq{eq:Hs} fragments
 \cite{motta2021}
\bea\label{LR_Ham}
\HH & =& \hat h_{\text{1e}}+\sum_{l=2}^{\Gamma}\HH_{l}^{(\text{LR})}, 
\eea
where
\bea
\hat h_{\text{1e}} &=& \HU_1^\dagger \left( \sum_p \epsilon_p^{(1)} \hn_p \right) \HU_1
=\sum_{pq}h_{pq}\CR{p}\AN{q}\\
\HH_{l}^{(\text{LR})} &=&  \HU_l^\dagger \left( \sum_{p,q} \epsilon_p^{(l)} \epsilon_q^{(l)} \hn_p \hn_q \right) \HU_l.
\eea

One computational advantage of this low-rank decomposition is that it can be done by diagonalizing 
the two-electron tensor $g_{pq,rs}$ considered as a matrix where each dimension is spanned by a pair of 
basis indices. This diagonalization gives a theoretical limit on $\Gamma \le N(N+1)/2$, where less sign
corresponds to a truncation of the expansion by removing terms for low magnitude eigenvalues.
Further details of this decomposition procedure can be found in Ref. \cite{motta2021}. 

 \emph{Full-rank (FR) optimization:} Using fragments of \eq{eq:Hmf} leads to the FR optimization (FRO) 
 \bea
H &= &  \hat h_{\text{1e}}+\sum_{l=2}^{\Gamma}\HH_{l}^{(\text{FR})},
\eea
where
\bea
\HH_{l}^{(\text{FR})} &=& \HU_l^\dagger \left(\sum_{i,j}^{N}\lambda_{ij}^{(l)}\hn_{i}\hn_{j} \right)\HU_l.\label{eq:CSA_dec}
\eea

 In practice, the set $\{\HH_l^{(\text{FR})}\}$
can be found by minimizing the $L_1$ norm of the $\mathbf{G}$ tensor
(under a given numerical threshold) in 
\bea\notag
\sum_{pqrs}^{N}g_{pqrs}\CR{p}\AN{q}\CR{r}\AN{s}&-&\sum_{l=2}^{\Gamma}\HU_{l}^{\dagger}(\mathbf{\theta}^{(l)})\left(\sum_{i,j}^{N}\lambda_{ij}^{(l)}\hn_i\hn_j\right)\HU_{l}(\mathbf{\theta}^{(l)})\\
&=&\sum_{prqs=1}^{N}G_{pqrs}\CR{p}\AN{q}\CR{r}\AN{s}\label{FRO_opt}
\eea
over the space of the variables $\{\mathbf{\theta}^{(l)}\}$
that parameterize the $\{\HU_{l}\}$ unitaries, as well as the
$\{\lambda_{ij}^{(l)}\}$ variables \cite{yen2021}.

A computationally more efficient variant of FR decomposition consists of an iterative
greedy strategy, which at the $i^{\rm th}$ iteration ($i\ge1$) finds the
single optimal Hamiltonian fragment $\HH_{i+1}^{(\text{FR})}$ that
minimizes the $L_1$ norm of tensor $\tilde{\mathbf{G}}^{(i+1)}$: 
\begin{equation}
\sum_{pqrs=1}^{N}\tilde{G}_{pqrs}^{(i)}\CR{p}\AN{q}\CR{r}\AN{s}-\HH_{i+1}^{(\text{FR})}=\sum_{prqs=1}^{N}\tilde{G}_{pqrs}^{(i+1)}\CR{p}\AN{q}\CR{r}\AN{s},\label{GFRO_opt}
\end{equation}
where $\tilde{G}_{pqrs}^{(1)}=g_{pqrs}$. Iterations are carried out
until the $L_1$ norm of $\tilde{\mathbf{G}}^{(i+1)}$ is below a given
threshold. The outlined scheme will be referred as greedy FR optimization (GFRO).
It yields larger norm Hamiltonian fragments in the beginning of the process and smaller norm ones 
at the end, which generally lowers the entropic part of \eq{Scal_Var} compared to other FR decomposition approaches.

{\it Initial Hamiltonian:} Historically, one-electron part of the electronic Hamiltonian was treated separately from 
the two-electron part \cite{motta2021,yen2021}. However, one can add the one electron contributions to the 
two-electron part by modifying the $g_{pq,rs}$ tensor. There are several ways to do this, one of the simplest 
approaches is to transform the entire Hamiltonian in the orbital frame where the one-electron part ($\hat h_{\text{1e}}$) is diagonal
\bea
\HU_1 \HH \HU_1^\dagger &=& \sum_p \epsilon_p\hn_p + \sum_{pq,rs} \tilde{g}_{pq,rs} \CR{p}\AN{q}\CR{r}\AN{s} \\ \notag
 \HH  &=& \HU_1^\dagger \left(\sum_{pq,rs} [\tilde{g}_{pq,rs} +\epsilon_p\delta_{pq}\delta_{pr}\delta_{ps}] \CR{p}\AN{q}\CR{r}\AN{s} \right) \HU_1\\ \label{eq:12eH}
 &=& \sum_{p'q',r's'} \bar{g}_{p'q',r's'} \CR{p'}\AN{q'}\CR{r'}\AN{s'},
\eea   
where primed indices correspond to orbitals after the $\HU_1$ conjugation. 
Using the augmented $\bar{g}_{p'q',r's'}$ tensor one can do either the LR or FR decompositions. 
In particular, we consider in this work GFRO for the Hamiltonian in \eq{eq:12eH}, 
which we termed SD-GFRO, where SD stands for "Singles and Doubles", as a reference to single and double excitation operators commonly used in electronic structure literature.

{\it Fragments post-processing:}  After obtaining solvable fragments $\{\HH_l\}$ one has a freedom to extract 
one-electron parts from each fragment 
\bea\notag
\HH_l &=& \HU_l^\dagger \left( \sum_{pq} \lambda^{(l)}_{pq} \hn_p \hn_q \right) \HU_l \\ \notag
&=& \HU_l^\dagger \left( \sum_{p} f^{(l)}_{p} \hn_p \right) \HU_l  \\
&+& \HU_l^\dagger \left( \sum_{pq} [\lambda^{(l)}_{pq}-\delta_{pq}f^{(l)}_{p}] \hn_p \hn_q \right) \HU_l,
\eea  
where the single index sum corresponds to the $p=q$ part and simplifies to a one-electron contribution 
due to $n_p^2=n_p$, and $f^{(l)}_{p}$ are free parameters that can be optimized. 
Here, $\lambda^{(l)}_{pq}$ can be either full-rank or rank-deficient depending on what 
scheme is used for the decomposition. 
This consideration allows one to sum all the one-electron parts from all fragments 
to obtain a new single solvable one-electron fragment 
\bea\notag
&&\hat h_{\text{1e}}+\sum_l \HU_l^\dagger \left( \sum_{p} f^{(l)}_{p} \hn_p \right) \HU_l = \\
&&\HU^\dagger \left( \sum_{p} \tilde{\lambda}_{pp} \hn_p \right) \HU.
\eea
The new Hamiltonian decomposition becomes 
\bea\notag
\HH &=&  \HU^\dagger \left( \sum_{p} \tilde{\lambda}_{pp} \hn_p \right) \HU \\
&+& \sum_l \HU_l^\dagger \left( \sum_{pq} [\lambda_{pq}^{(l)}-\delta_{pq}f^{(l)}_{p}] \hn_p \hn_q \right) \HU_l.
\eea  

A possible advantage of this regrouping can come 
from potential cancellation of contributions within the single one-electron term so that the spectral range of the 
new one-electron term is reduced and reduction of spectral norms for two-electron terms after removing the 
diagonal parts. Optimizing parameters $f^{(l)}_p$ was performed successfully for the measurement problem in the 
VQE framework \cite{ChoiFFF2022}, yet similar optimization for the Trotter error runs into computational challenges 
associated with accurate prediction of fragment spectral range dependence on $f^{(l)}_p$. Due to this difficulty 
we employed a heuristic approach that is based on the connection between a fragment spectral range and  
an $L_1$ norm for its LCU decomposition (see \eq{eq:L1UB}). 

The idea of the heuristic approach is to use fragments whose LCU decomposition has low $L_1$ norm, such fragments 
can be obtained with a particular choice of $f^{(l)}_p$. It was found in Ref. \cite{lee2021}
that substitution of every $\hn_i$ operator in two-electron parts by reflection $\hr_{i}=(1-2\hn_{i})$ ($\hr_i^2 = 1$ and 
$\hr_i^\dagger = \hr_i$) reduces the $L_1$ norm of a collection of solvable fragments. 
This substitution will require adjustment of coefficients and addition of one electron terms 
\bea\notag
    H &=& \sum^{N}_{pq=1}\tilde{h}_{pq}a^{\dagger}_{p}\AN{q} \\ \notag
    &+&\sum_{k=2} \HU_k^{\dagger}\left(\sum_{i, j}^{N}\frac{\lambda_{ij}^{(k)}}{4}\hr_{i}\hr_{j} \right) \HU_{k} \\
    &-&\frac{1}{4}\sum_{p,q} g_{pp,qq},\label{LCUHam}
\eea 
where
\begin{equation}\label{RenormHopp}
    \tilde{h}_{pq}=h_{pq}+\sum_{k}g_{pq,kk}.
\end{equation}
This procedure reduces four-fold the two-body tensor within the Hamiltonian fragments which generally leads to a decrease of their respective spectral norms. 
{\it Circuit analysis:}
To estimate the number of T-gates we provide the estimate for the number of single-qubit rotation
gates for all fermionic Hamiltonian partition schemes considered here.
The implementation of the Hamiltonian (first-order) Trotter step is 
\begin{equation}
\prod_{n=1}^{\Gamma} e^{-i \HH_n t} =\hat{U}^{\dagger}_{1}e^{-i\sum_{p}\epsilon^{(1)}_{p}\hat{n}_{p}}\hat{U}_{1}
\prod_{l=2}^{\Gamma}\HU_l^\dagger e^{-i t\sum_{i,j}^{N}\lambda_{ij}^{(l)}\hn_{i}\hn_{j}} \HU_l \label{eq:Trot_LR}
\end{equation}
where adjacent unitary spin-orbital rotations can be combined using the Lie group closure
$\HU_l \HU_{l+1}^\dagger=\HU_{l,l+1}$, $\HU_{l,l+1}$ has the form of \eq{eq:Uor} with $N(N-1)/2$ one-electron generators.  
Each of these $\HU_{l,l+1}$ rotations, can
be decomposed as a set of $N(N-1)/2$ rotations that contain 
2 arbitrary amplitude single-qubit rotations and Clifford transformations \cite{kivlichan2020,motta2021}.
Similarly, the circuit implementation
for the simulation of the $\exp\left(-i t\sum_{i,j}^{N}\lambda_{ij}^{(l)}\hn_{i}\hn_{j}\right)$
(note that $\lambda_{ij}^{(l)}=\epsilon_{i}^{(l)}\epsilon_{j}^{(l)}$ for LR) is
accomplished with at most $N(N+1)/2$ two-qubit gates, 
where the compilation of each one of the latter entails the implementation of 2 single-qubit rotations. Finally, the implementation of $\exp\left(-i\sum_p\epsilon^{(1)}_{p}\hat{n}_{p}\right)$ requires at most $N$ single qubit rotations.
From these considerations and taking into account that a single Trotter step requires the implementation of $\Gamma+1$ orbital rotations, it follows that the total single rotation gate
count is no larger than $2N^{2}\Gamma-N$. 

\subsection{Qubit partitioning methods}

Qubit realizations of the electronic Hamiltonian are obtained after applying
one of the fermion-qubit mappings \cite{Jordan1928,bravyi2002fermionic} 
\begin{equation}
\HH_{q}=\sum_{n}c_{n}\hat{P}_{n},\quad\hat{P}_{n}=\otimes_{k=1}^{N}\hat{\sigma}_{k}^{(n)}\label{eq:Qub_Ham}
\end{equation}
where $c_{n}$ are numerical coefficients and $\hat{P}_{n}$ are tensor products of single-qubit Pauli operators
and the identity, $\hat{\sigma}_{k}^{(n)}=\hat{x}_{k},\hat{y}_{k},\hat{z}_{k},\hat I_{k}$, acting on the $k$th qubit.

\emph{Fully-commuting (FC) grouping:} This approach partitions $\HH_q$ into $\HH_n^{(\text{FC})}$
fragments containing commuting Pauli products: if $\hat{P}_{I},\hat{P}_{J}\in \HH_n^{(\text{FC})}$
then $[\hat{P}_{I},\hat{P}_{J}]=0$. This commutativity condition guarantees that 
$\HH_n^{(\text{FC})}$ can be rotated into a linear combination of $\hat z_i$ operator products 
by a sequence of Clifford group transformations \cite{yen2020measuring, Zack2022}.
We will be referring to it as the fully-commuting qubit partitioning scheme, mainly because it was developed 
for the VQE measurement problem where ``fully" was added to its name to separate it from a more restrictive 
qubit-wise commuting scheme \cite{verteletskyi2020}. 

The problem of finding the minimum number of fragments $\{\HH_n^{(\text{FC})}\}$ representing 
 the Hamiltonian $\HH_q$ has been shown to be equivalent to a Minimum
Clique Cover (MCC) problem for a graph representing the Hamiltonian. For this graph 
each term of $\HH_q$ is a vertex and the commutativity condition determines connectivity \cite{yen2020measuring,verteletskyi2020}.
The MCC problem is NP-hard but it was found that different heuristic polynomial-in-time algorithms
can find good approximate solutions. In this work we specifically
consider the largest-first (LF) heuristic, whose name makes a reference
to the ordering which the algorithm uses to process the vertices of
the Hamiltonian graph \cite{yen2020measuring,verteletskyi2020}; as well as the Sorted
Insertion (SI) algorithm \cite{crawford2021}. The latter is a greedy algorithm introduced
to reduce the number of measurements required to attain a given level
of accuracy in the estimation of the energy expectation value (or
any other observable), rather than aiming at the minimization of the
number of groups in the Hamiltonian partition.

{\it Circuit analysis:}
Here we analyze the number of single-qubit rotations as a proxy for the number of T-gates.  
For both qubit commutativity grouping schemes, the first order Trotter propagator can be implemented as   
\bea
\prod_{n=1}^{\Gamma} e^{-i t\HH_n}
= \prod_{n=1}^{\Gamma} [\HU_n^{\dagger} e^{-i t \hat Z_n} \HU_n ] \label{eq:Trotter_QWC}
\eea
where $\hat Z_{n}$ is a linear combination of Pauli products of $\hat z_i$ operators, and 
$\HU_n$ are single- and two-qubit Clifford transformations for the FC scheme.
Due to the Clifford character, $\HU_n$'s do not contribute to the T-gate count.  
The number of $e^{-i t \hat Z_{n}}$ terms scale as  $\mathcal{O}(N^{3})$ while the total number of 
single-qubit rotations will be the same as the total number of Pauli products in the Hamiltonian, $\mathcal{O}(N^{4})$ \cite{yen2020measuring}.

For a summary of the Hamiltonian decomposition methods considered in this work, we refer the reader to Table \ref{summaryMeths}.
\begin{table*}
\centering
\begin{tabular}{|c|c|c|c|c|}
\hline 
Method & Partition Encoding & Algorithm character & Extra-processing & Refs.\tabularnewline
\hline 
\hline 
FC-LF & Qubit & Non-greedy & - &\cite{verteletskyi2020} \tabularnewline
\hline 
FC-SI & Qubit & Greedy & - & \cite{verteletskyi2020},\cite{yen2020measuring},\cite{van2020} \tabularnewline
\hline 
LR & Fermionic & - & - &\cite{motta2021} \tabularnewline
\hline 
LR-LCU & Fermionic & - & Post-processing & \cite{Ignacio_LCU:2022}\tabularnewline
\hline 
FRO & Fermionic & Non-greedy & - & \cite{yen2021}\tabularnewline
\hline 
GFRO & Fermionic & Greedy & - & \cite{yen2021}\tabularnewline
\hline 
GFRO-LCU & Fermionic & Greedy & Post-processing & \cite{Ignacio_LCU:2022}\tabularnewline
\hline 
SD-GFRO & Fermionic & Greedy & Pre-processing & \cite{Ignacio_LCU:2022} \tabularnewline
\hline 
\end{tabular}

\caption{Summary of the Hamiltonian decomposition methods considered in this
work, alongside the type of encoding employed in the partition (fermionic
or qubit-based), the character of the algorithm (greedy, non-greedy)
and whether the scheme includes extra-processing of the Hamiltonian
previous or after its decomposition.}\label{summaryMeths}

\end{table*}

\subsection{Use of symmetries for Trotter error estimation}

We can exploit symmetries shared by all Hamiltonian fragments pertaining
to the same partition to introduce tighter estimations of the first-order Trotter
error. In practice, Hamiltonian
simulation can be performed on an initial state belonging to one of
the symmetry group irreducible representations \cite{su2021}. Therefore, $\alpha $
can be estimated by considering only the subspace of the Hilbert space corresponding to 
the irreducible representation of interest.

Here, we extended the approach developed in Ref. \cite{su2021},
where estimations of Trotter error are based on fermionic norms
for particle-number preserving operators $\{\hat X\}$ ($[\hat X,\hat{N}]=0$,
$\hat{N}$ is the total-particle number operator). These norms are
equal to the spectral norm of the observables defined on manifolds
spanned by eigenstates of the total particle-number operator ($\hat{N}|\psi_{\eta}\rangle=\eta|\psi_{\eta}\rangle$):
\begin{equation}
||\hat X||_{\eta}=\underset{|\psi_{\eta}\rangle}{\text{max}}\sqrt{\langle\psi_{\eta}|\hat X^{\dagger}\hat X|\psi_{\eta}\rangle}.
\end{equation}

It is straightforward to generalize this metric to include the spin
symmetries $\hat S_{z}$ (projection of the total electronic spin angular
momentum along $z$) and $\hat S^{2}$ (the square of the norm of the total
electronic spin angular momentum) for the fermionic partition approaches.
Consequently, the projected Trotter error is performed on manifolds
that are spanned by states ${|\psi_{\eta,m,s}\rangle}$ which are
simultaneous eigenstates of $\hat{N}$, $\hat S_{z}$ and $\hat S^{2}$, with
eigenvalues $\eta$, $m$ and $s(s+1)$, respectively. Similarly,
for qubit-based methods, it is possible to find qubit symmetries $\{\hat{Q}_{i}\}$
that satisfy $[\hat{Q}_{i},\HH_n]=0$ $\forall n$
{[}see Eq. (\ref{Partition}){]}. A general search for these qubit symmetries can be computationally 
expensive. Therefore we restrict our consideration only to a single Pauli product symmetries, 
they can be found efficiently using techniques developed for the qubit tapering \cite{bravyi1701tapering}.

 For the Trotter analysis we introduce 
\begin{equation}
\alpha _{\mathbf{Q}}=\sum_{n_{1},n_{2}}||[\HH_{n_{2}},\HH_{n_{1}}]||_{\mathbf{Q}}.\label{eq:alpha_Q}
\end{equation}
where $\mathbf{Q}$ labels the set of quantum numbers that define
the manifold the Trotter error is projected on. Thus, for fermionic
methods we have $\mathbf{Q}=[n,m,s]$ whereas for qubit-based schemes
$\mathbf{Q}=[\zeta_{1},\zeta_{2},\dots,\zeta_{n}]$, where $\zeta_{i}$
labels one of the eigenvalues of the $i$th qubit symmetry $\hat{Q}_{i}$.
Symmetry constraints are straightforwardly extended to the $\beta$ spectral range 
upper bound for $\alpha$ 
\bea
\beta_{\mathbf{Q}}=\Delta E_{i,\mathbf{Q}}\Delta E_{j,\mathbf{Q}}\ge\alpha _{\mathbf{Q}},\label{eq:beta_Q}
\eea
where $\Delta E_{i,\mathbf{Q}}$ is the spectral range
of the symmetric manifold with quantum numbers $\mathbf{Q}$ for $\hat H_i$.

\section{Results and discussion}

\label{Results}

\begin{figure*}[ht!]
\centering \includegraphics[scale=0.8]{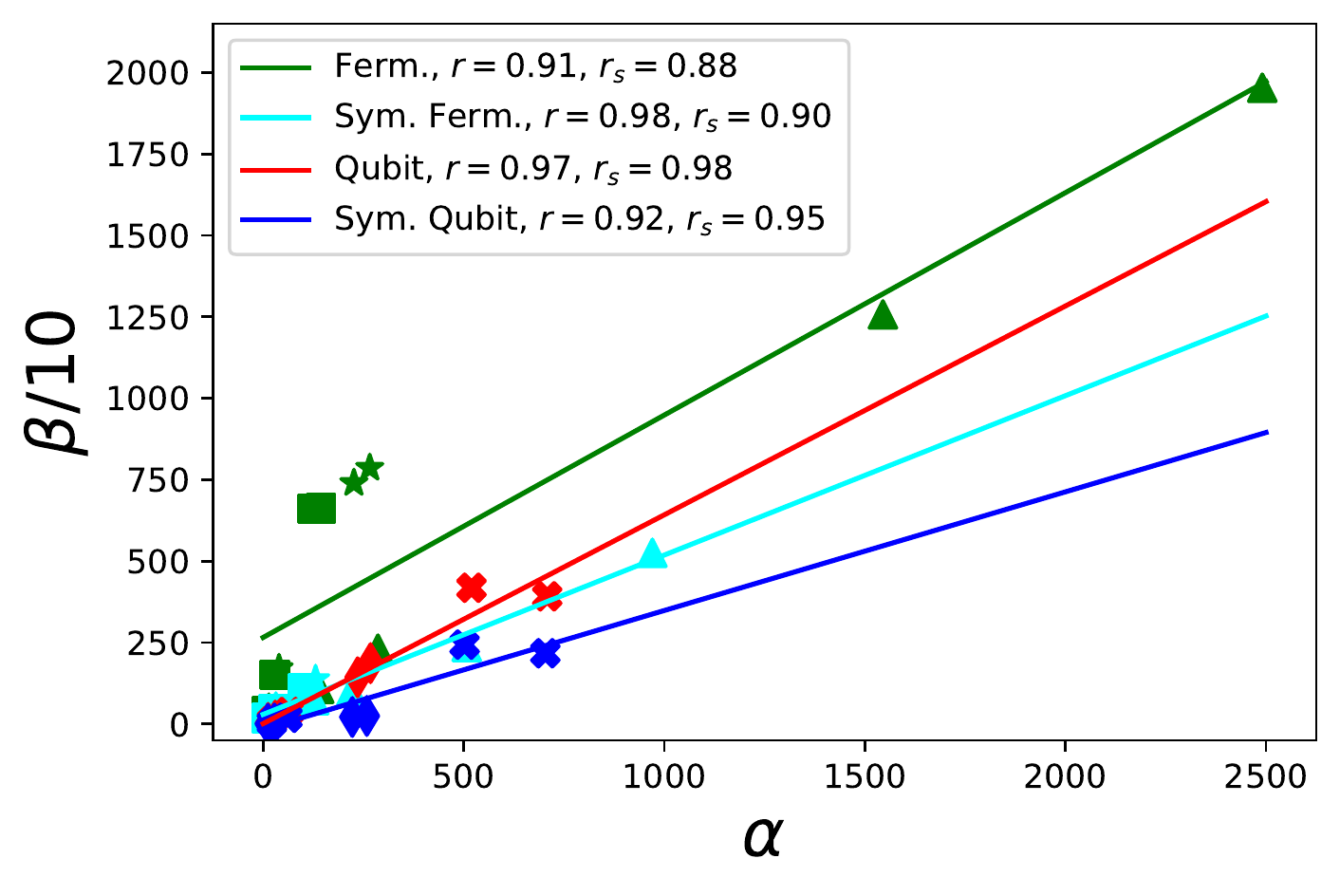}
\caption{Correlations between $\alpha $ and its upper bound $\beta$
{[}Eqs. (\ref{alphaP}) and (\ref{TightBet}){]} for fermionic (green) and qubit (red) methods. The correlation between the symmetry-projected counterparts $\alpha_{\mathbf{Q}}$ and $\beta_{\mathbf{Q}}$ [see Eqs. (\ref{eq:alpha_Q}) and (\ref{eq:beta_Q})] are shown in cyan and blue for fermionic and qubit methods, respectively. The fermionic methods considered are FRO (triangles), GFRO (squares) and LR (stars), whereas the qubit partition methods are FC LF (crosses) and FC SI (diamonds).
The straight lines are obtained by a least-square fit, their Pearson ($r$) and Spearman ($r_{s}$) coefficients are indicated in the legend.}
\label{fig:Estalpha}
\end{figure*}

We considered the partition schemes described in the Methods section
for the electronic Hamiltonians corresponding to molecules H$_{2}$,
LiH, BeH$_{2}$, H$_{2}$O and NH$_{3}$. The latter were generated
using the STO-3G basis and the Bravyi-Kitaev transformations (for
the qubit encodings) as implemented in the OpenFermion package \cite{mcclean2020openfermion}.
The nuclear geometries for the molecules are given by R(H-H)=1 {\AA}
(H$_{2}$), R(Li-H)=1 {\AA} (LiH) and R(Be-H)=1 {\AA} with collinear
atomic arrangement (BeH$_{2}$), R(O-H)=1 {\AA} and $\angle$HOH=107.6$^{\circ}$
(H$_{2}$O); and R(N-H)=1 {\AA} with $\angle$HNH=107$^{\circ}$ (NH$_{3}$).

\begin{figure*}[ht!]
\centering \includegraphics[scale=0.6]{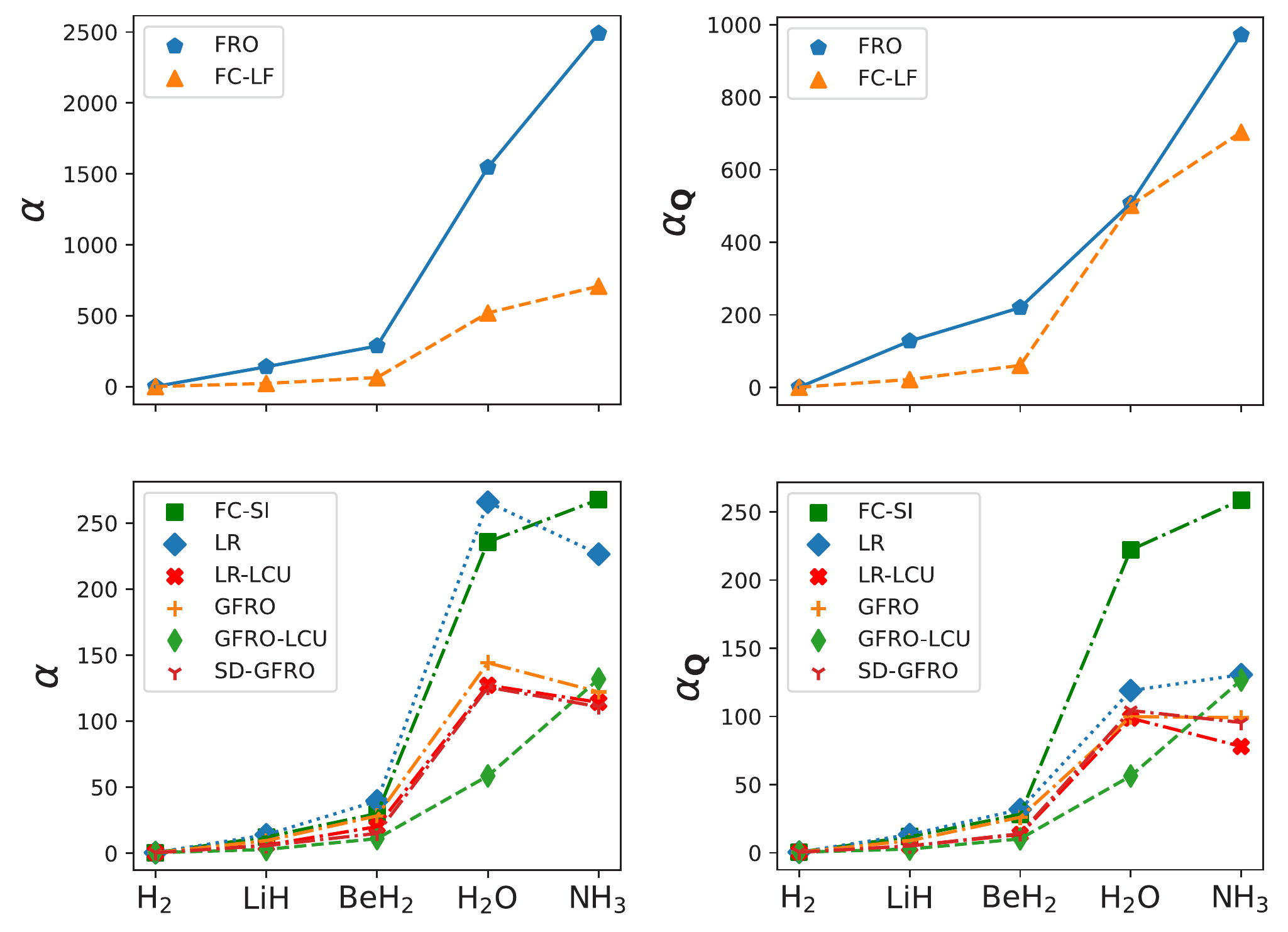}
\caption{Left panels: trends in Trotter errors $\alpha$ (in squared Hartrees)
for the different Hamiltonian partitionings. Right panels: trends in symmetry-projected Trotter errors $\alpha_{\mathbf{Q}}$ (in squared Hartrees). For the fermionic methods we consider $\alpha_{\eta,m,s}$ {[}see Eq. (\ref{eq:alpha_Q}){]}
with $\eta$ corresponding to the number of electrons of the neutral
molecule, and $m=s=0$. For the qubit-based methods, we
consider $\alpha_{\mathbf{Q}}$
{[}see Eq. (\ref{eq:alpha_Q}){]} in the subspace of qubit symmetries corresponding to the ground state of the neutral molecule.For clarity and to highlight the difference in the scale of the non-greedy algorithms with respect to the rest of the methods, we show FRO and FC-LF results in different plots. }
\label{Summ_alpha}
\end{figure*}

\subsection{Trotter errors}
Here we explore the relationship between Trotter errors with 1) the greedy/non-greedy character of the Hamiltonian decomposition techniques and 2) pre- and post-processing of Hamiltonian fragments. For that end, we mainly rely on the upper bounds $\beta$ and descriptors $C$, $S_{L}$ introduced in subsection \ref{TrotterErrorEst}.\\

First, we demonstrate the quality of our spectral range based upper-bounds, $\beta$ and $\beta_{\mathbf{Q}}$, 
for Trotter error estimates, $\alpha$ and $\alpha_{\mathbf{Q}}$. Fig.~\ref{fig:Estalpha} shows that even though 
$\beta$'s still yield somewhat loose upper bounds of $\alpha$'s, the two quantities are well correlated.
Imposing symmetry constraints, generally, makes correlation better and obviously lowers the values of $\alpha$ and $\beta$. 
Good correlations between $\alpha$ and $\beta$ will allow us to analyze results for different partitioning methods 
for Trotter errors (Fig.~\ref{Summ_alpha}) in terms of spectral distributions of fragments.

\begin{figure*}[ht!]
\centering \includegraphics[scale=0.6]{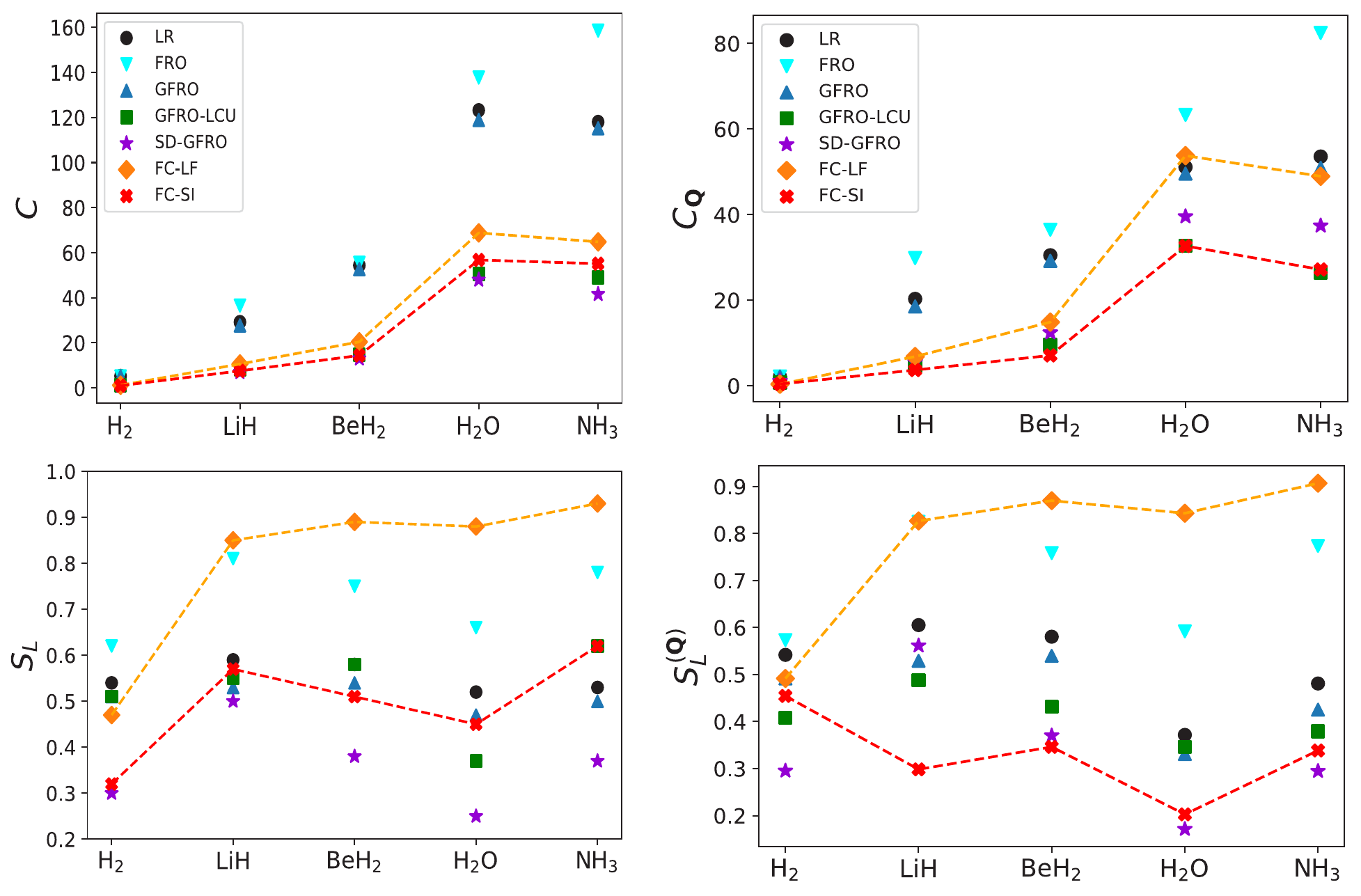}
\caption{Left panels: $C$ and $S_{L}$ [Eqs. (\ref{LinEntropy}) and (\ref{CandS})] descriptors of spectral range distribution of Hamiltonian fragments for different partition methods. As a visual aid, descriptor data associated to qubit methods is highlighted with dashed lines. Right panels: symmetry-projected versions of  $C$ and $S_{L}$, $C_{\mathbf{Q}}$ and $S^{(\mathbf{Q})}_{L}$, where the subindex $\mathbf{Q}$ denotes the set of quantum numbers that label the symmetric manifold of the projection.}
\label{NormsFerm-1}
\end{figure*}

From Fig.~\ref{Summ_alpha}, it is evident that Trotter errors
are consistently smaller for the greedy approaches compared to the non-greedy counterparts for both
fermionic and qubit techniques (cf. FRO with GFRO, and FC-LF with FC-SI). Using greedy approaches and LCU-inspired post processing as well as one- and 
two-electron combining SD-pre-processing make fermionic techniques more accurate than their qubit counterparts.
Further insights on these observations can be drawn with the aid of Fig.~\ref{NormsFerm-1} , where the $S_{L}$ and $C$ descriptors for the different methods are presented.  
We can use the latter to understand the large $\alpha$'s corresponding to water and ammonia partitioned with the FRO scheme, and the relatively close Trotter errors for the rest of molecules which lead to a clustering in Fig.~\ref{fig:Estalpha}. The reasons are 1) the larger spectral range of fragments obtained for these molecules and method (see Fig.~\ref{NormsFerm-1}) and 2) the rapid (quadratic) dependence of the error with the sum of these ranges as can be seen in Eq. (\ref{Scal_Var}).

We see from Fig.~\ref{NormsFerm-1}  a significant difference in magnitudes of $\beta$ depending on whether the method is fermionic- or qubit-based, but the correlated and monotonic $\alpha-\beta$ relationship is maintained within each family.
Therefore, the descriptors $S_{L}$ and $C$ introduced in Eq. (\ref{CandS}) should be compared between methods pertaining to the same decomposition technique type (fermionic or qubit-based). 

However, it is interesting to note that qubit methods tend to yield smaller $C$ values with respect to most of the fermionic methods. We conjecture that this is due to the less restrictive optimization of fragments compared to fermionic methods, for which spin and particle number symmetries are preserved, whereas that is not the case for qubit methods. In fact, after mapping fermionic Hamiltonian fragments to qubit encodings we notice that the number of Pauli words tend to be significantly larger than the average number of Pauli words of qubit Hamiltonian fragments. These additional number of Pauli words, appear to account for the conservation of symmetries for fermionic Hamiltonian fragments, which ultimately can lead to larger spectral ranges (and $C$ metrics) for the latter.
We note from Fig.~\ref{NormsFerm-1} that greedy algorithms, GFRO, SD-GFRO, and FC-SI exhibit both lower $C$ and $S_{L}$ 
values with respect to the non-greedy versions (cf. FC-LF with FC-SI). This observation implies that Hamiltonian decomposition methods that favor low and non-uniformly distributed spectral ranges in the resulting Hamiltonian fragments tend to reduce the Trotter error.

On top of the greedy character of Hamiltonian fragments, we show in Fig.~\ref{Summ_alpha} that the LCU post-processing can 
introduce further improvements in Trotter error. From Fig.~\ref{NormsFerm-1} it is evident that postprocessing endows smaller spectral ranges to the resulting Hamiltonian fragments, reflected in the decrease of both $C$ and $C_{\mathbf{Q}}$. 

The only molecule that has not benefited from this procedure is NH$_{3}$ for which LCU-post processing increases Trotter error (Fig.~\ref{Summ_alpha}). This is in contrast with the descriptors $S_{L},C$ and its symmetry-projected versions $C_{\mathbf{Q}},S^{(\mathbf{Q})}_{L}$, that predict a lower Trotter error after LCU post-processing. We attribute this trend violation to the loose character of the descriptors to estimate $\alpha$ and $\alpha_{\mathbf{Q}}$, which fail to quantitatively predict changes in Trotter error after post-processing. Note that tighter Trotter error estimators would be needed especially for NH$_{3}$ where the relative change in Trotter error is smaller than the rest of molecules for post-processing. 

Interestingly, according to Fig.~\ref{NormsFerm-1}, SD-GFRO is expected to be the best performing fermionic method in terms of Trotter error when considering the non-symmetry projected $C$ and $S_{L}$ descriptors. This is mainly due to the non-uniform
spectral range distribution for its Hamiltonian fragments. However, when considering symmetry-projected descriptors, which provide a tighter estimation of the Trotter error, we note that GFRO-LCU becomes competitive, as the latter introduces in some Hamiltonian decompositions either fragments with smaller spectral norms, or a more uneven spectral range distributions with respect to those in SD-GFRO. 
In fact, similar Trotter errors are observed for both methods, as summarized in Fig.~\ref{Summ_alpha}.

\subsection{Number of single-qubit rotations}
\begin{figure*}[ht!]
\centering \includegraphics[scale=0.6]{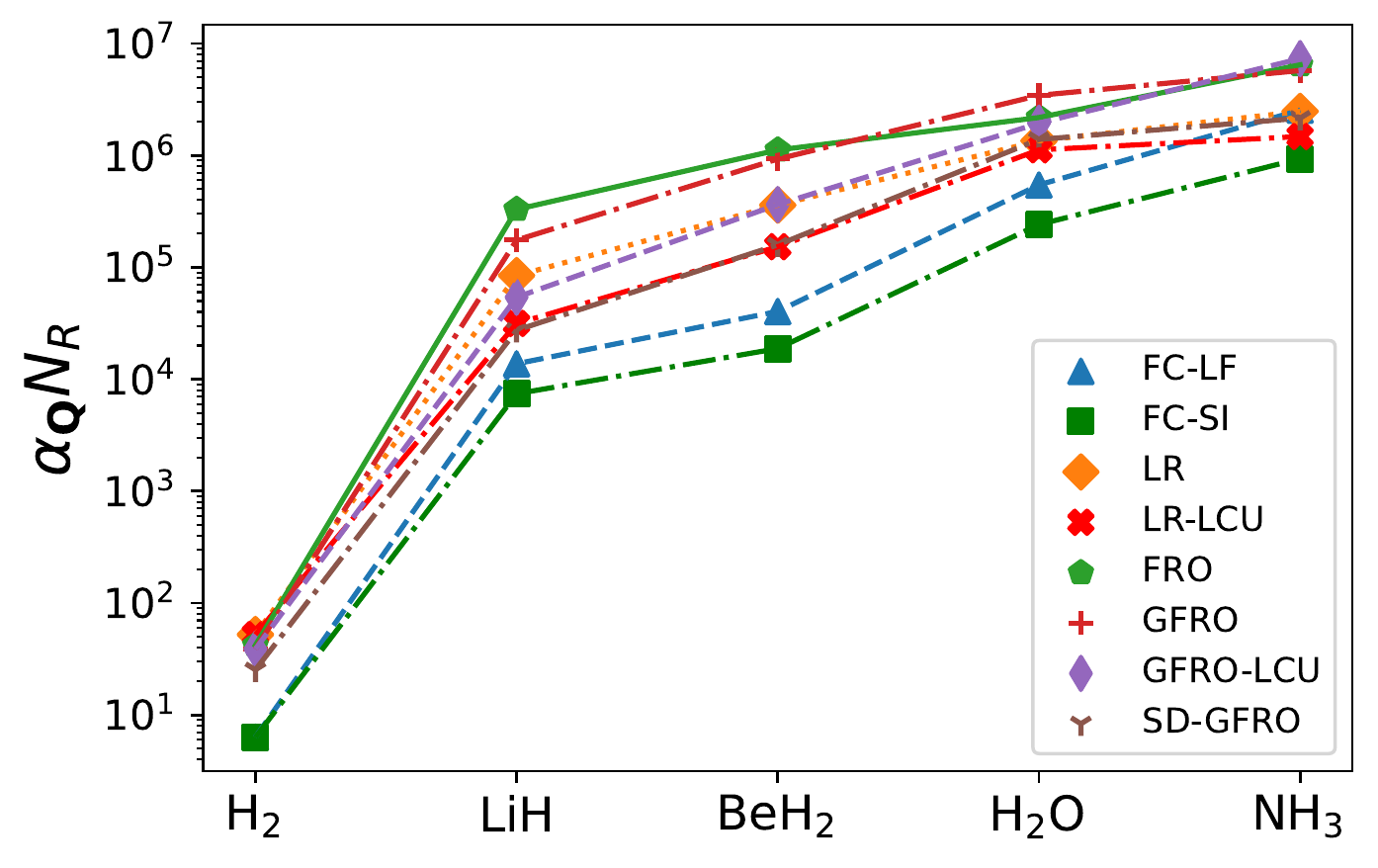}
\caption{Symmetry-projected Trotter errors (in squared Hartrees) scaled by the number of single qubit rotations in each Trotter step, $N_{R}$, for the Hamiltonian partitions. As highlighted in the main text, this figure of merit gauges the complexity of the circuits required for energy eigenvalue estimation for a Trotterized (Adaptive) QPE algorithm.}
\label{Tgate_costs}
\end{figure*} 

\begin{table*}[t]
\centering
\begin{tabular}{|c|c|c|c|}
\hline 
Molecule & 1st Best ($N_{T}$) & 2nd Best ($N_{T}$) & 3rd Best ($N_{T}$)\tabularnewline
\hline 
\hline 
H$_{2}$ & FC-SI (9.45$\times$10$^{8}$) & FC-LF (9.45$\times$10$^{8}$) & SD-GFRO (4.33$\times$10$^{9}$)\tabularnewline
\hline 
LiH & FC-SI (1.94$\times$10$^{12}$) & FC-LF (3.73$\times$10$^{12}$) & SD-GFRO (7.73$\times$10$^{12}$)\tabularnewline
\hline 
BeH$_{2}$ & FC-SI (5.19$\times$10$^{12}$) & FC-LF (1.16$\times$10$^{13}$) & LR-LCU ($4.73\times10^{13}$)\tabularnewline
\hline 
H$_{2}$O & FC-SI ($7.6\times10^{13}$) & FC-LF ($1.79\times10^{14}$) & LR-LCU ($3.82\times10^{14}$)\tabularnewline
\hline 
NH$_{3}$ & FC-SI ($3.16\times10^{14}$) & LR-LCU ($5.09\times10^{14}$) & SD-GFRO ($7.56\times10^{14}$)\tabularnewline
\hline 
\end{tabular}
\caption{Best resource-efficient Hamiltonian decomposition methods for the
tested molecules for eigenvalue estimation within a $10^{-3}$ h error
with a Trotterized Adaptive QPE algorithm. An upper bound estimation of T-gate
count $N_{T}$ is indicated with parenthesis. The latter are calculated according to the procedure outlined in Appendix \ref{NumTgat}.}\label{TgateUBs}
\end{table*}

Figure \ref{Tgate_costs} shows the results for our metric of complexity of circuits required for eigenvalue estimation under a Trotterized Adaptive QPE scheme.
Symmetry-projected $\alpha_{\mathbf{Q}}$ is used for a tighter estimation of circuit complexity.
Even though fermionic methods have smaller $\alpha_{\mathbf{Q}}$ and $\Gamma$ compared to the qubit-based counterparts, 
the single rotation gate count $N_{R}$ is lower for the greedy qubit techniques which ultimately yields more resource-efficient circuits for the latter. This can be attributed to two factors: First, the (upper bound for) number of single-qubit 
rotations for qubit techniques is the same as the number of terms in the Hamiltonian ($\mathcal{O}(N^{4})$), 
and it does not change with the partitioning. Second, for fermionic fragments the number of single qubit rotation count per fragment scales $\mathcal{O}(N^{2})$, but the number of fragments grows as $\mathcal{O}(N^{2})$ or faster in greedy versions. In addition, 
prefactors of these dependencies play a decisive role for fixed systems. 

Finally, we highlight that in spite of its simplicity, our figure of merit $\alpha_{\mathbf{Q}}N_{R}$ is a good indicator of resource efficiency, as it predicts the best performing methods when we carry out upper-bound T-gate counts ($N_{T}$) that take into account T-gate synthesis approximations and limited phase estimation resolution as sources of error in eigenvalue estimation (see Fig. \ref{Tgate_costs} and Table \ref{TgateUBs} ). For the range of the $\alpha_{\mathbf{Q}}N_{R}$ metric considered in this work, its fidelity for resource estimation is due to the monotonic character of $N_{T}$ with the former (see Appendix \ref{NumTgat}). It is an open question whether this monotonic dependence is preserved for larger $\alpha_{\mathbf{Q}}N_{R}$ and it is out of the scope of the present work.

\section{Conclusions}
\label{Conclusions}

We have assessed first-order Trotter product formula errors for different partitioning schemes of electronic Hamiltonians. 

Taking advantage of Hamiltonian fragment symmetries improves the tightness of 
Trotter error estimates if the propagated wavefunction is symmetric. 
Greedy algorithms yield lower Trotter errors than their non-greedy counterparts, which can be rationalized by introducing 
the upper bound for the Trotter error that uses spectral ranges of individual fragments.
This upper bound allowed us to identify two main error contributions: 1) sum over fragment Hamiltonian spectral ranges and
 2) linearized entropy factor that favors uneven distributions and explains success of greedy techniques. 

These observations are expected to hold for higher-order Trotter formulas,
for which it has been proved that the commutator-dependent scaling
of Trotter error up to leading order in the simulation time
step is \cite{childs2021Trot}: 
\begin{equation}
\alpha^{(p)}=\sum_{n_{1},n_{2},\dots,n_{p+1}}^{\Gamma}||[\HH_{n_{p+1}},\cdots[\HH_{n_{2}},\HH_{n_{1}}]\cdots]||
\end{equation}
$p$ being the order of the Trotter formula. For instance, one can write the upper bound for the second-order Trotter error as 
\begin{equation}
\alpha^{(2)} \le 2\sum_{n_{1},n_{2},n_{3}}^{\Gamma}||\HH_{n_{1}}||\cdot||[\HH_{n_{2}},\HH_{n_{3}}] ||.
\end{equation}

To obtain the upper bound for the right hand side we can use the same ideas as in \eq{Scal_Var}, which give an estimator 
for the second order Trotter error
\begin{equation}
\alpha^{(2)}\le \frac{C^{3} S_{L}}{2}.
\end{equation} 

Hence, we conjecture that decompositions of the Hamiltonian that favors low spectral range Hamiltonian fragments as well as non-uniformity in the distribution of spectral ranges, such as greedy algorithms, are beneficial to the purpose of Trotter error lowering, even beyond the first order Trotter.

On top of the greedy strategy, LCU post-processing of Hamiltonian fermionic fragments can improve Trotter errors even more. 
However, fermionic decompositions are not the most resource-efficient for Hamiltonian simulation
considering the number of single qubit rotations, which is proportional to the number of T-gates expected after compilation into a quantum circuit. Qubit-based decompositions have lower numbers of single-qubit rotations compared to fermionic ones. 

We note that scaling of the number of single-qubit rotations as a function of simulated spin-orbitals (qubits) between qubit methods and the most resource-efficient fermionic partitioning ({\it i.e.} LR-LCU) is the same, and it goes as $\mathcal{O}(N^{4})$. 
It is an interesting question to explore whether qubit techniques remain as the most economical in the large 
$N$ limit. For that end, a better understanding of the scaling of the number of greedy fermionic fragments 
(which afford the lowest Trotter errors) with the number of simulated spin orbitals as well as the implementation 
of truncation in the number of fragment schemes, is subject of future study.

\section*{Acknowledgments}

Authors thank Ignacio Loaiza and Priyanka Mukhopadhyay for useful discussions. 
L.A.M.M is grateful to the Center for Quantum Information and Quantum Control (CQIQC) for a postdoctoral fellowship. A.F.I.
acknowledges financial support from the Google Quantum Research Program, 
and Zapata Computing Inc. This research was
enabled in part by support provided by Compute Ontario and Compute
Canada.

\appendix
\section{T-gate count estimations.}\label{NumTgat}
 Here, we elaborate on upper-bound estimations for T-gate count for a fixed target error $\epsilon$ in energy eigenvalue estimation in a Trotterized Adaptive Quantum Phase Estimation algorithm \cite{berry2009,kivlichan2020}. The total T-gate count $N_{T}$, as pointed out in previous works \cite{reiher2017,kivlichan2020} is given by 
 \begin{equation}
  N_{T}=N_{R}N_{HT}N_{PE}
 \end{equation}
where $N_{R}$ is the number of single-qubit rotations needed for the implementation of a single Trotter step in a quantum computer. $N_{HT}$ refers to the number of T gates needed to compile one single qubit rotation (for a fixed target error $\epsilon_{HT}$) and $N_{PE}$ is the number of Trotter steps required to resolve the target energy eigenvalue under a target uncertainty $\epsilon_{PE}$, the latter scaling as $t^{-1}$, $t$ being the total simulation time. The Trotter errors $\alpha, \alpha_{\mathbf{Q}}$ can be shown \cite{kivlichan2020}  to define the scaling of an upper bound in the energy eigenvalue estimation error incurred by the Trotter approximation of unitary evolution, and is given by $\epsilon_{T}=\alpha t$. 
$N_{T}$ can be written in terms of the aforementioned errors as
\begin{equation}
N_{T}	\approx\frac{0.76\pi\alpha N_{R}}{|\epsilon_{T}|\epsilon_{PE}}\left[1.15\log_{2}\left(\frac{N_{R}\alpha}{\epsilon_{HT}|\epsilon_{T}|}\right)+9.2\right]
\end{equation}
 By restricting the target error in eigenvalue estimation to $\epsilon$, we have that $\epsilon=\epsilon_{T}+\epsilon_{HT}+\epsilon_{PE}$. It is generally the case \cite{kivlichan2020}  that $\epsilon_{T}>\epsilon_{PE}\gg\epsilon_{HT}$, such that
\begin{equation}
\begin{split}
 N_{T}  & \approx\frac{0.76\pi\alpha N_{R}}{|\epsilon_{T}|(\epsilon-|\epsilon_{T}|)} \\
 \times & \left[1.15\log_{2}\left(\frac{N_{R}\alpha}{(\epsilon-\epsilon_{PE}-|\epsilon_{T}|)|\epsilon_{T}|}\right)+9.2\right].
 \end{split}
\end{equation} 
Thus, we can optimize over the target errors $|\epsilon_{T}|,\epsilon_{PE}$ to minimize the number of T-gates $N_{T}$.
We performed calculations of $N_{T}$ for all methods and molecules addressed in this work and the best performing Hamiltonian decomposition techniques are summarized in Table \ref{TgateUBs}.
The reason behind our choice of gauging the complexity of the circuits for each partition method by means of the figure of merit $\alpha_{\mathbf{Q}} N_{R}$ instead of $N_{T}$ (Fig.~\ref{Tgate_costs}), is the simplicity of the former and highlights the relevance of both the error $\alpha_{\mathbf{Q}}$ as well as the number of Hamiltonian fragments encoded in $N_{R}$ that ensues from the decomposition methods. Indeed this figure of merit exhibits high fidelity in comparing the complexity of circuits among decomposition methods as outlined in the main text.
 
\section{Computational details.}
The Hamiltonian fragments for molecules and methods considered in this work, alongside the code used for the calculation of Trotter errors and its analysis can be accessed in a developer version of TEQUILA platform \cite{kottmann2021tequila} available at https://github.com/lamq317/tequila .
For reproduction of Trotter error data, see the script https://github.com/lamq317/tequila/blob/pr-troterr/tests/testTrotErr.py .
All Hamiltonian fragments, except for SD-GFRO, were generated using a code based on the Scipy optimize library \cite{2020SciPy-NMeth} and the BFGS algorithm \cite{fletcher2013} for fermionic decomposition methods. SD-GFRO fragments were generated using the code available at https://github.com/iloaiza/MAMBO . The latter can also be used to generate the rest of Hamiltonian fragments considered in this work, using different numerical libraries.

\clearpage{} \bibliographystyle{quantum}
\bibliography{TrotterBench,Trotter}

\end{document}